\documentclass[10pt,twocolumn,letterpaper]{article}

\usepackage[pagenumbers]{cvpr} 

\usepackage{graphicx}
\usepackage{amsmath}
\usepackage{amssymb}
\usepackage{booktabs}

\usepackage{epsfig}
\usepackage{cite}
\usepackage{verbatim}
\usepackage{color}
\usepackage{array}
\usepackage{epstopdf}
\usepackage{booktabs, multirow}       
\usepackage{stackengine}
\usepackage{colortbl}
\usepackage{multirow}
\usepackage{makecell}
\usepackage{algorithm}
\usepackage{algpseudocode}

\newcommand{\bPy}{\ensuremath{{\mathbf{P_y}}}}
\newcommand{\bPyy}{\ensuremath{{\mathbf{P}_\mathbf{y}^2}}}

\usepackage[pagebackref,breaklinks,colorlinks]{hyperref}

\usepackage[capitalize]{cleveref}
\crefname{section}{Sec.}{Secs.}
\Crefname{section}{Section}{Sections}
\Crefname{table}{Table}{Tables}
\crefname{table}{Tab.}{Tabs.}

\def\cvprPaperID{*****} 
\def\confName{CVPR}
\def\confYear{2022}

\begin{document}

\title{RD-Optimized Trit-Plane Coding of Deep Compressed Image Latent Tensors}

\author{Seungmin Jeon\\
Korea University\\
{\tt\small seungminjeon@mcl.korea.ac.kr}
\and
Jae-Han Lee\\
Korea University\\
{\tt\small jaehanlee@mcl.korea.ac.kr}
\and
Chang-Su Kim\\
Korea University\\
{\tt\small changsukim@korea.ac.kr}
}
\maketitle

\begin{abstract}
DPICT is the first learning-based image codec supporting fine granular scalability. In this paper, we describe how to implement two key components of DPICT efficiently: trit-plane slicing and rate-distortion-optimized (RD-optimized) coding. In DPICT, we transform an image into a latent tensor, represent the tensor in ternary digits (trits), and encode the trits in the decreasing order of significance. For entropy encoding, it is necessary to compute the probability of each trit, which demands high time complexity in both the encoder and the decoder. To reduce the complexity, we develop a parallel computing scheme for the probabilities, which is described in detail with pseudo-codes. Moreover, we compare the trit-plane slicing in DPICT with the alternative bit-plane slicing. Experimental results show that the time complexity is reduced significantly by the parallel computing and that the trit-plane slicing provides better RD performances than the bit-plane slicing.
\end{abstract}

\section{Introduction}
\label{sec:intro}

Image compression is a fundamental topic in image processing.
There are classical codecs, including JPEG \cite{codec_jpeg}, JPEG2000 \cite{codec_jpeg2000} and BPG \cite{bpg}, which achieve efficient data compression and transmission based on hand-crafted modules, such as DCT and wavelets.
Recently, with its great success in various fields of image processing, deep learning has been adopted for image compression as well.

Among learning-based codecs, convolutional neural network (CNN) codecs \cite{y2017_ICLR_balle, y2018_ICLR_balle, y2018_NIPS_minnen, y2020_CVPR_cheng} exhibit competitive rate-distortion (RD) performances, as compared to the classical codecs.
They transform an image into latent representation using an encoder network and then compress it into a bitstream.
Then, the latent representation is inversely transformed into a reconstructed image by a decoder network.
Ball{\'e} \textit{et al}.\ \cite{y2017_ICLR_balle, y2018_ICLR_balle} proposed adding a uniform noise to overcome the non-differentiability of quantization in the training phase and using an autoencoder for hyper-priors.
Minnen \textit{et al}.\ \cite{y2018_NIPS_minnen} adopted context models to compress latent representation more compactly.
Cheng \textit{et al}.\ \cite{y2020_CVPR_cheng} assumed Gaussian mixture models for latent representation.
However, these CNN codecs support fixed-rate compression only, so the networks should be trained for $N$ times to support $N$ different rates.

\begin{figure}[t]
    \begin{center}
    \includegraphics[width=\linewidth]{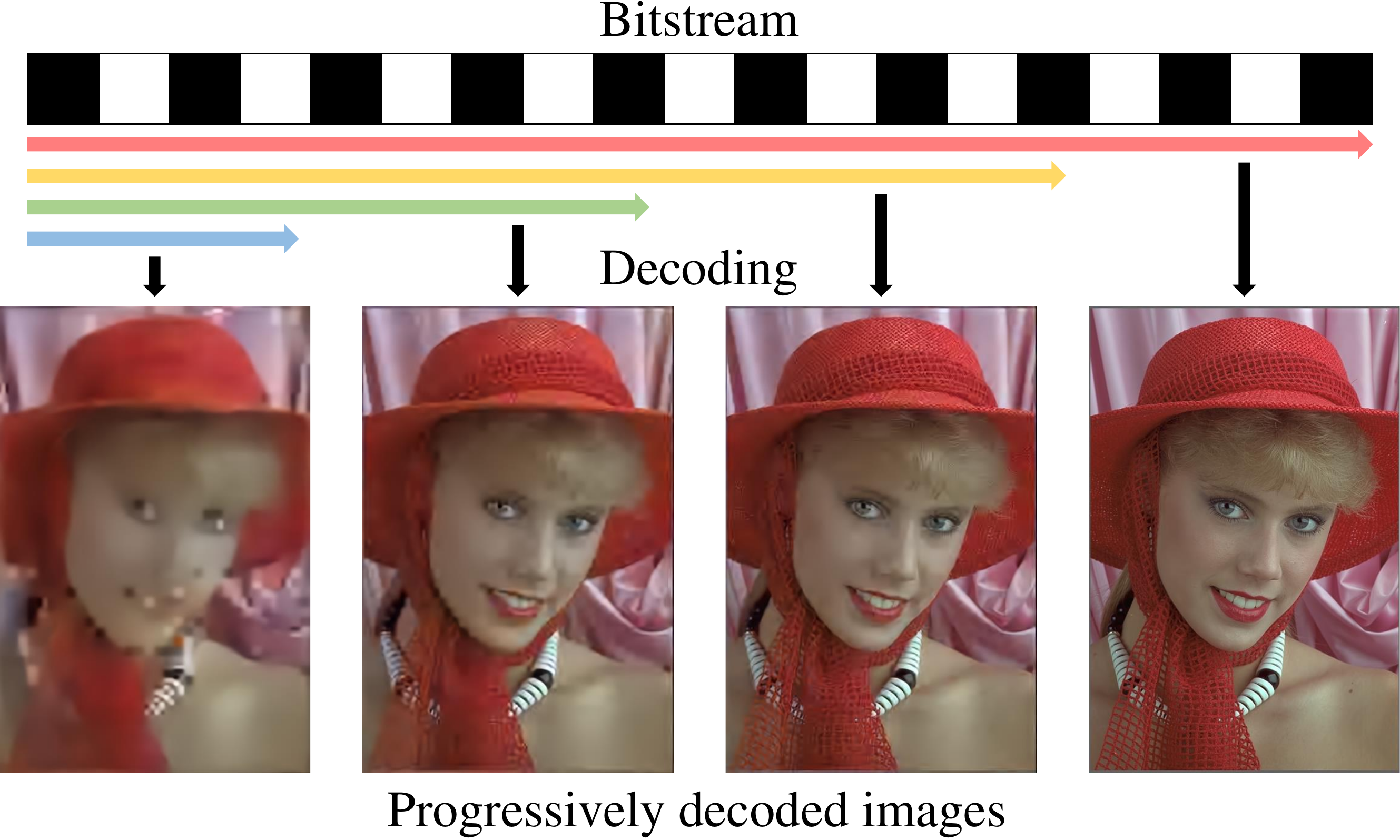}
    \end{center}
    \vspace*{-0.2cm}
    \caption
    {
        Illustration of progressive image compression.
    }
    \label{fig:progressive_coding_abstract}
\end{figure}

Progressive compression, or scalable coding \cite{y2005_IEEE_ohm}, is useful in applications, for it encodes an image into a single bitstream that can be reconstructed at various levels of quality at different bit-rates, as illustrated in Figure~\ref{fig:progressive_coding_abstract}. There are many display devices with different sizes and resolutions, and network conditions also vary greatly. By employing a progressive image codec, a single bitstream can be truncated according to network conditions and then transmitted to multiple decoders to be displayed on different devices. Thus, progressive compression is more efficient than non-progressive one.

Some learning-based codecs \cite{y2016_ICLR_toderici, y2016_NIPS_gregor,y2017_CVPR_toderici,y2018_CVPR_johnston}, as well as classical JPEG and JPEG2000,  support progressive compression, most of which are based on recurrent neural networks (RNNs). Toderici \textit{et al}.\ \cite{y2016_ICLR_toderici} proposed the first RNN for progressive compression, which transmits a bitstream progressively using long short-term memory (LSTM). Gregor \textit{et al}.\ \cite{y2016_NIPS_gregor} introduced a recurrent codec to improve conceptual image quality using a generative model. However, these codecs \cite{y2016_ICLR_toderici, y2016_NIPS_gregor} are for low-resolution patches only. Toderici \textit{et al}.\ \cite{y2017_CVPR_toderici} developed a codec for higher-resolution images by expanding the work in \cite{y2016_ICLR_toderici}. Johnston \textit{et al}.\ \cite{y2018_CVPR_johnston} proposed an effective initializer for hidden states of their RNN and a spatially adaptive rate controller. However, the performances of these RNN-based codecs are inferior even to those of traditional codecs.

DPICT \cite{y2022_CVPR_DPICT} is the first CNN-based progressive image codec to support fine granular scalability (FGS). In other words, its bitstream can be truncated at any point to reconstruct the scene faithfully. In this paper, we describe in detail how to implement the DPICT algorithm, which performs trit-plane slicing and RD-optimized coding of trits. A naive implementation of these components would take unnecessarily long encoding and decoding times, for the probability mass functions (PMFs) of all trits in a latent tensor should be estimated. To reduce the time complexity, we develop parallel computing schemes for the trit-plane slicing and the RD-optimized coding. Also, we discuss the effectiveness of the trit-plane slicing in DPICT in comparison with the alternative bit-plane slicing. Experiments demonstrate that the parallel computing reduces the time complexity significantly and that the trit-plane slicing yields better RD performances than the bit-plane slicing.

\vspace*{0.5cm}
\section{Algorithm}
\label{sec:algorithm}
Let us analyze the two key components of DPICT: trit-plane slicing and RD-optimized coding. For comparison, we also describe the bit-plane slicing, which can replace the trit-plane slicing.

\begin{figure}[t]
    \begin{center}
    \includegraphics[width=\linewidth]{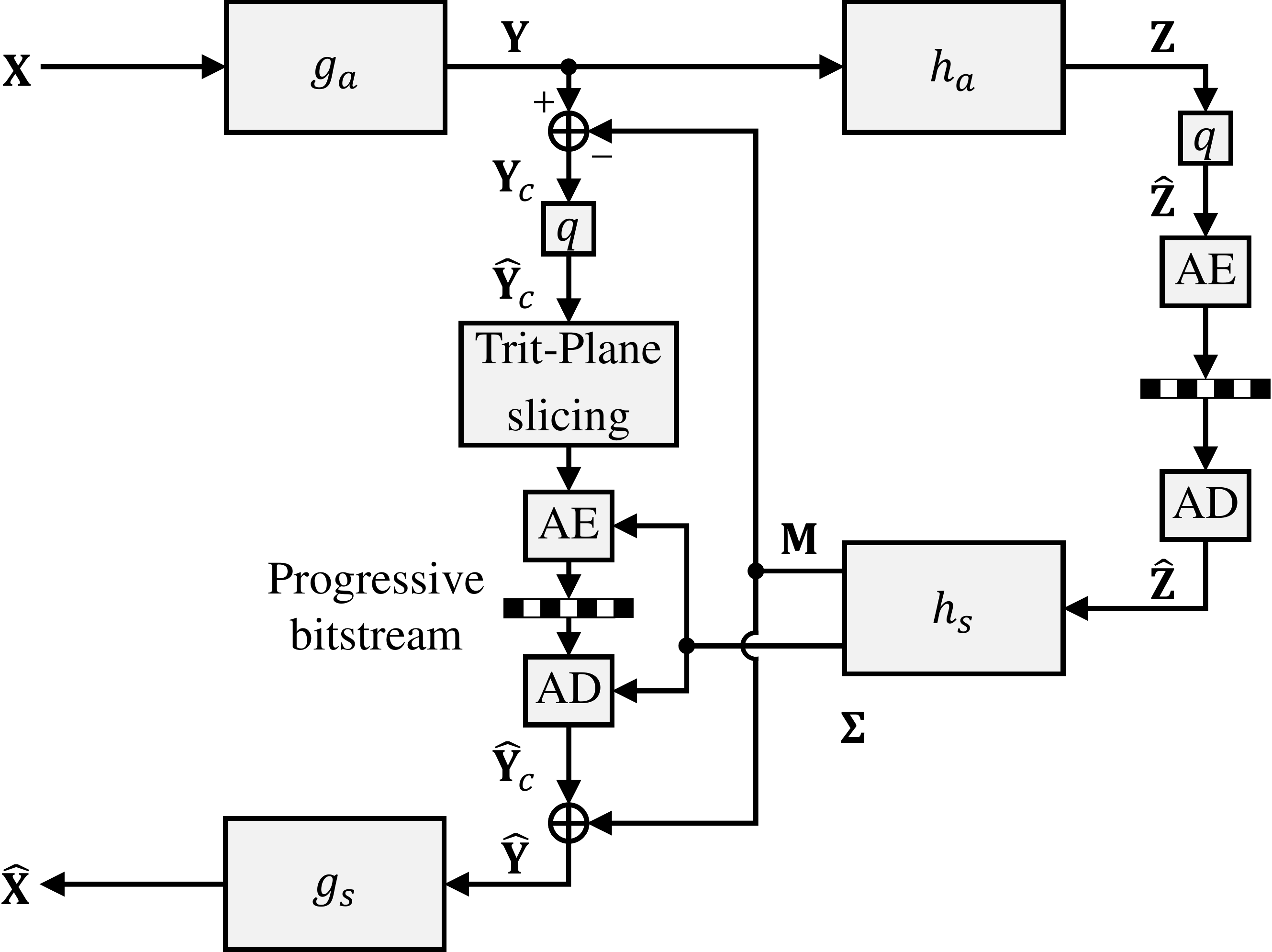}
    \end{center}
    \vspace*{-0.2cm}
    \caption
    {
        An overview of the DPICT network. AE and AD mean arithmetic encoder and decoder, respectively.
    }
    \label{fig:network_architecture}
\end{figure}

Figure~\ref{fig:network_architecture} is an overview of DPICT, where $\mathbf{X}$ is an RGB image and $\mathbf{Y} \in \mathbb{R}^{C \times H \times W}$ is a latent tensor generated by the encoder. Also, $\mathbf{M} \in \mathbb{R}^{C \times H \times W}$ and $\mathbf{\Sigma} \in \mathbb{R}^{C \times H \times W}$ represent the means and standard deviations of latent elements in $\mathbf{Y}$, respectively.
$\mathbf{Y}_c$ is the mean-removed $\mathbf{Y}$, and $\hat{\mathbf{Y}}_c$ is a quantized version of $\mathbf{Y}_c$. More specifically,
\begin{equation}
\label{eq:quantization}
\hat{\mathbf{Y}}_c=q(\mathbf{Y}_c)=q(\mathbf{Y}-\mathbf{M})
\end{equation}
where rounding is used for the quantization function $q$. Then, $\hat{\mathbf{Y}}_c$ is entropy-encoded into a progressive bitstream. Note that we adopt the same network structure as Cheng \textit{et al}.~\cite{y2020_CVPR_cheng} do, but we omit the autoregressive model for estimating entropy parameters \cite{y2018_NIPS_minnen,y2020_CVPR_cheng}. In this work, the latent tensor is reconstructed progressively plane-wise in an RD-optimal manner, instead of element-wise in the raster scan order. Therefore, the autoregressive model, assuming perfect reconstruction of causal elements, cannot be used.

\vspace*{0.3cm}
\subsection{Bit-plane and trit-plane slicing}
\label{ssec:bit_and_trit_plane_slicing}
\subsubsection{Bit-plane slicing}
\label{sssec:bit_plane_slicing}

We can represent $\hat{\mathbf{Y}}_c$ in the binary number system and encode it based on the bit-plane slicing.

Let $y_i$ be the $i$th element in $\hat{\mathbf{Y}}_c$ and $\sigma_i$ be its standard deviation. Here, $0\leq i \leq (K-1)$, and $K=C \times H \times W$ is the number of elements in $\hat{\mathbf{Y}}_c$. Although $y_i$ may have an infinite value theoretically, its interval $\textstyle {\cal I}_0$ is clipped finitely by
\begin{equation}
\label{eq:I_0}
{\cal I}_0 = [l_0, r_0) = [ \textstyle -\frac{2^l - 1}{2}, \frac{2^l + 1}{2})
\end{equation}
where
\begin{equation}
\label{eq:tail}
l = \left\lceil{\log_2(2{\cdot}{\sigma_i}{\cdot}\Phi^{-1}(1-\epsilon))}\right\rceil.
\end{equation}
$\Phi(\cdot)$ denotes the cumulative density function (CDF) of the standard normal distribution, and $\epsilon$ is a tiny value set to $5\times10^{-10}$. Thus, $l$ is about $6.1 + \log_2 \sigma_i$. Note that ${\cal I}_0$ is determined to be the smallest interval of the length of a power of 2, over which the integral sum of distribution is larger than $1-\epsilon$.

For the bit-plane slicing, we divide ${\cal I}_0$ into two subintervals of the same length. Let $b_0$ denote the first bit of $y_i$. We set $b_0=0_{(2)}$ if $y_i < {\textstyle \frac{l_0+r_0}{2}}$, and $b_0=1_{(2)}$ otherwise. This process is repeated. More specifically, let ${\cal I}_n=[l_n, r_n)$ be the interval where $y_i$ belongs when the first $n$ bits $\{b_{n-1}, \ldots, b_{0}\}$ are encoded. Then, ${\cal I}_n$ is divided into two subintervals:
\begin{align}
\label{eq:interval_example_1}
\textstyle
{\cal I}_n^0 = [l_n^0, r_n^0) = [l_n, \textstyle \frac{l_n+r_n}{2}),\\
\label{eq:interval_example_2}
{\cal I}_n^1 = [l_n^1, r_n^1) = [\textstyle \frac{l_n+r_n}{2}, r_n).
\end{align}
Then, the next bit $b_{n}$ is determined according to the subinterval that $y_i$ belongs to. The corresponding subinterval becomes ${\cal I}_{n+1}$.

The 1s place bits of the whole elements in $\hat{\mathbf{Y}}_c$ form the least significant bit-plane, the 2s place bits form the next bit-plane, and so on. The arithmetic encoder \cite{y2013_arXiv_duda_ANS} compresses these bit-planes sequentially from the most significant bit-plane to the least significant one. For the arithmetic coding, both the encoder and the decoder should compute the conditional PMF of $b_n$ given the information $b_{n-1}, \cdots ,b_0$,
\begin{align}
\label{eq:cond_prob}
P(b_{n}=k|b_{n-1}, & \cdots, b_0) = P(y_i \in {\cal I}_{n}^k | y_i \in {\cal I}_{n}) \\
& =  \frac{\Phi(r_{n}^k/\sigma_i) - \Phi(l_{n}^k/\sigma_i)}{\Phi(r_{n}/\sigma_i) - \Phi(l_{n}/\sigma_i)}
\end{align}
where $k \in \{0_{(2)}, 1_{(2)}\}$.

At the decoder, we reconstruct each latent element using partially decoded bit-planes. Suppose that the first $n$ bits $\{ b_{n-1},\cdots,b_0 \}$ are available for $y_i$. Then, we reconstruct $y_i$ to
\begin{equation}
\label{eq:recon_of_y}
\textstyle
\hat{y}_i^n = E[y_i|y_i \in {\cal I}_n],
\end{equation}
which minimizes the mean squared error
\begin{equation}\label{eq:distortion}
\textstyle
D_n = E[(y_i - \hat{y}_i^n)^2|y_i \in {\cal I}_n].
\end{equation}
In other words, $\hat{y}_i^n$ in \eqref{eq:recon_of_y} is the minimum mean square error (MMSE) estimate.

\subsubsection{Trit-plane slicing}
\label{sssec:trit_plane_slicing}
\begin{figure}[t]
    \begin{center}
    \includegraphics[width=\linewidth]{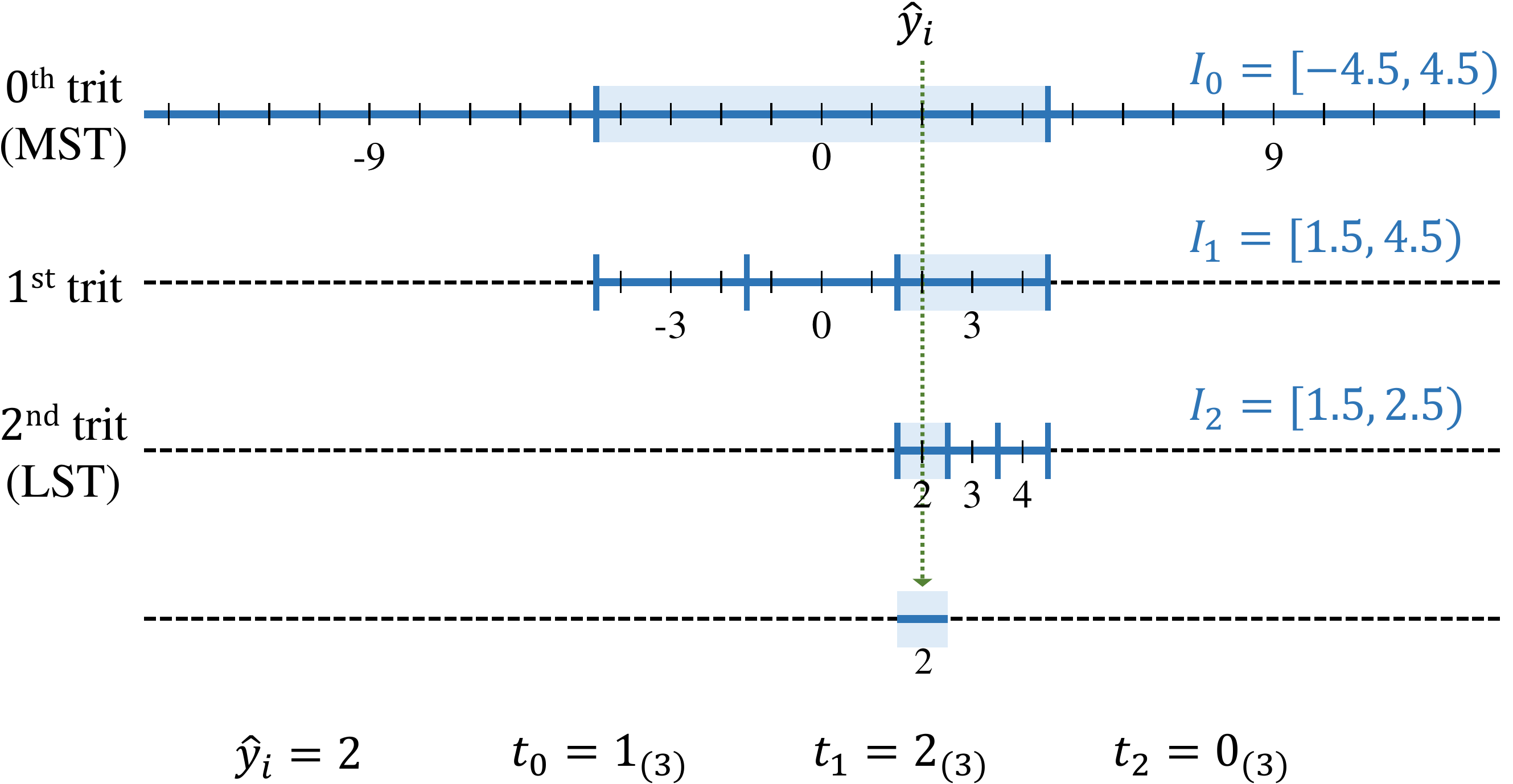}
    \end{center}
    \vspace*{-0.2cm}
    \caption
    {
        Illustration of the trit-plane slicing.
    }
    \label{fig:trit_plane_slicing}
\end{figure}

The trit-plane slicing in DPICT is similar to the bit-plane slicing, but it divides an interval into three subintervals \cite{y2022_CVPR_DPICT}, instead of two. First, instead of (\ref{eq:I_0}), the interval for $y_i$ is defined as
\begin{equation}
\label{I_0_trit}
\textstyle
{\cal I}_0 = [l_0, r_0) = [\textstyle -\frac{3^l - 1}{2}, \textstyle \frac{3^l + 1}{2}).
\end{equation}
Also, an interval ${\cal I}_n$ is divided into three subintervals:
\begin{align}
\textstyle
{\cal I}_{n}^0 &= [l_n^0, r_n^0) = [l_n, \textstyle \frac{2l_n + r_n}{3}),  \\
{\cal I}_{n}^1 &= [l_n^1, r_n^1) = [\textstyle \frac{2l_n + r_n}{3}, \textstyle \frac{l_n + 2r_n}{3}), \\
{\cal I}_{n}^2 &= [l_n^2, r_n^2) = [\textstyle \frac{l_n + 2r_n}{3}, r_n).
\end{align}
Using these subintervals, the $n$th trit $t_{n} \in \{0_{(3)}, 1_{(3)}, 2_{(3)}\}$ is determined similarly to Section \ref{sssec:bit_plane_slicing}. Figure \ref{fig:trit_plane_slicing} shows an example of the trit-plane slicing. The $3^0$s place trits form the least significant trit-plane (LST), the $3^1$s place trits form the next trit-plane, and so on. These trit-planes are compressed by the arithmetic coder \cite{y2013_arXiv_duda_ANS}. The reconstructed $\hat{y}_i^n$ and the corresponding distortion $D_n$ are computed by \eqref{eq:recon_of_y} and \eqref{eq:distortion}, respectively.

\vspace*{0.3cm}
\subsection{RD-optimized coding}\label{ssec:rd_prioritized_transmission}
To transmit more important information first, we compress trit-planes
from the most significant trit-plane (MST) to LST. Moreover, within each trit-plane, we transmit the $K$ trits after sorting them according to their RD priorities. The transmission of a trit increases the rate ($\Delta R > 0$) but decreases the distortion ($\Delta D < 0$). The goal is to minimize $\Delta R$ and maximize $-\Delta D$ simultaneously. To achieve this goal, sophisticated RD methods \cite{y1998_SPM_ortega} can be applied. However, for simplicity, we use the ratio $- \frac{\Delta D}{\Delta R}$ for the RD-optimized coding.

Suppose that the $n$th trit-plane is to be compressed. Thus, for an element $\hat{y}_c$, its previous trits $\{ t_{n-1},\cdots,t_0 \}$ are already transmitted and its $n$th trit $t_n$ is to be compressed. Both the encoder and the decoder can compute the probabilities
\begin{equation}
q_k = P(t_n=k | t_{n-1}, \cdots, t_0),
\end{equation}
$k \in \{ 0_{(3)}, 1_{(3)}, 2_{(3)} \}$, similarly to \eqref{eq:cond_prob}. Then, the expected number of bits for encoding $t_n$ is given by the entropy $H(\{q_0, q_1, q_2\})$,
\begin{equation}
\textstyle
\Delta R = H(\{q_0, q_1, q_2\}) = - \sum_{k=0}^{2} q_k \log_2 q_k.
\label{eq:DeltaR}
\end{equation}
Also, for the three cases of $t_n = 0_{(3)}, 1_{(3)}, 2_{(3)}$, we compute the distortions via \eqref{eq:distortion}, respectively, which are denoted by $D_{n+1}^0, D_{n+1}^1, D_{n+1}^2$.
Then, the expected distortion change is given by
\begin{equation}
\textstyle
\Delta D = E[D_{n+1}]-D_{n}= \sum_{k=0}^{2} q_k D_{n+1}^k - D_n
\label{eq:DeltaD}
\end{equation}
Finally, we compute the RD priority of the $n$th trit, which is defined as
\begin{equation}
\label{eq:priority}
- \frac{\Delta D}{\Delta R} = \frac{\sum_{k=0}^{2} q_k D_{n+1}^k - D_{n}}{\sum_{k=0}^{2} q_k \log_2 q_k}.
\end{equation}
Then, we transmit the $K$ trits in the trit-plane in the decreasing order of their RD priorities. In this way, more important information is transmitted first, even in the same trit-plane \cite{y2022_CVPR_DPICT}.

\vspace*{0.5cm}
\section{Implementation Details}

Let us detail how to implement the trit-plane slicing and the RD-optimized coding and make them perform efficiently. For simplicity, we omit the bit-plane slicing in this section, which can be implemented similarly to the trit-plane slicing.

A latent tensor is composed of many trits, each of which has its own PMF. As described in Section~\ref{ssec:bit_and_trit_plane_slicing}, the PMFs of these trits should be computed recursively. These recursive computations are costly, but they are essential for the RD-optimized coding. By performing these computations parallel using matrices in $\mathtt{Pytorch}$ or $\mathtt{NumPy}$, we can reduce the computation time. Algorithms \ref{alg:plane_slicing} and \ref{alg:approach2} are pseudo-codes for parallel computing the trit-plane slicing and the RD-optimized coding, respectively. The full source codes are available at \cite{y2021_github_lee_DPICT}.

\begin{algorithm}[t]
\caption{Trit-plane slicing}\label{alg:plane_slicing}
    {\bf Input:} Mean-removed, quantized latent tensor $\hat{\mathbf{Y}}_c$ and standard deviation tensor $\mathbf{\Sigma}$
    \begin{algorithmic}[1]
        \State{$\mathbf{L} \gets \lceil{\log_3(2{\cdot}{\mathbf{\Sigma}}{\cdot}\Phi^{-1}(1-\epsilon))}\rceil$ \Comment{$\epsilon=5\times10^{-10}$}}
        \State{$\mathbf{S} \gets {\hat{\mathbf{Y}}_c}+\lfloor {3 ^ \mathbf{L}} \div 2 \rfloor$}
        \State{$L_{\max} \gets \max(\mathbf{L})$}
        \ForAll{$n = 1, 2, \ldots, L_{\max}$}
            \State{$\mathbf{T}_{n-1}$ $\gets$ $\lfloor \mathbf{S} \div 3^{L_{\max}-n} \rfloor$ }
            \State{$\mathbf{S} \gets \mathbf{S} \mod 3^{L_{\max}-n}$}
        \EndFor
    \end{algorithmic}
    {\bf Output:} Trit-planes $\mathbf{T}_0, \mathbf{T}_1, \ldots,  \mathbf{T}_{L_{\max}-1}$
\end{algorithm}

\begin{algorithm}[b]
\caption{RD-optimized coding (naive implementation)}\label{alg:approach1}
    {\bf Input:} Trit-planes $\mathbf{T}_0, \mathbf{T}_1, \ldots,  \mathbf{T}_{L_{\max}-1}$
    \begin{algorithmic}[1]
        \ForAll{${\mathbf{T}_n}, n=0, \ldots, L_{\max}-1$}
            \ForAll{trits $t_n$ in $\mathbf{T}_n \in \mathbb{R}^{C \times H \times W}$}
            \State{Compute $P(t_n=k|t_{n-1}, \ldots ,t_0)$ \Comment{as in \eqref{eq:cond_prob}}}
            \State{$\Delta R \gets - \sum_{k=0}^{2} q_k \log_2 q_k$ \Comment{ (\ref{eq:DeltaR})}}
            \State{$\Delta D \gets \sum_{k=0}^{2} q_k D_n^k - D_{n-1}$ \Comment{ (\ref{eq:DeltaD})}}
            \State{Save the priority $-\Delta D / \Delta R$}
            \EndFor
            \State{Sort the trits in the decreasing order of priorities}
            \State{Entropy-encode the sorted trits into the bitstream}
        \EndFor
    \end{algorithmic}
    {\bf Output:} Compressed bitstream
\end{algorithm}

In Algorithm \ref{alg:plane_slicing}, we transform each element in a latent tensor into a series of trits to form trit-planes. First, in line 1, we obtain the tensor $\mathbf{L}$ for the lengths of the intervals ${\cal I}_0$ for all elements, which is the parallel implementation of the trit-plane equivalent of \eqref{eq:tail}. Then, we determine the interval of each element similarly to \eqref{eq:I_0}. The interval is zero-centered, so it is shifted to the right to start at 0. In line 2, $\mathbf{S}$ is the tensor for such shifted intervals. In lines 4$\sim$7, by dividing $\mathbf{S}$ by powers of 3, we obtain the trit-planes $\mathbf{T}_0, \mathbf{T}_1, \ldots, \mathbf{T}_{L_{\max}-1}$. Note that $\mathbf{T}_0$ is the MST, while $\mathbf{T}_{L_{\max}-1}$ is the LST. In fact, they are not `planes' literally, because each $\mathbf{T}_n \in \mathbb{R}^{C \times H \times W}$ is a tensor of the same size as $\hat{\mathbf{Y}}_c$.

Algorithm \ref{alg:approach1} summarizes a naive, non-parallel implementation of the RD-optimized coding, which computes the RD priority of each trit in a trit-plane individually using a nested for-loop. For the conditional PMF computations in line 3 and the entropy coding in line 10, we use functions in the $\mathtt{CompressAI}$ library \cite{y2020_arXiv_begaint_compressai}. Also, note that line 3 is performed similarly to \eqref{eq:cond_prob}. As described in Section~\ref{ssec:rd_prioritized_transmission}, we compute the RD priority of each trit in lines 3$\sim$6. This naive approach, however, is time-consuming, for it requires as many iterations as the number of all trits.

\begin{figure}[t]
    \begin{center}
    \includegraphics[width=\linewidth]{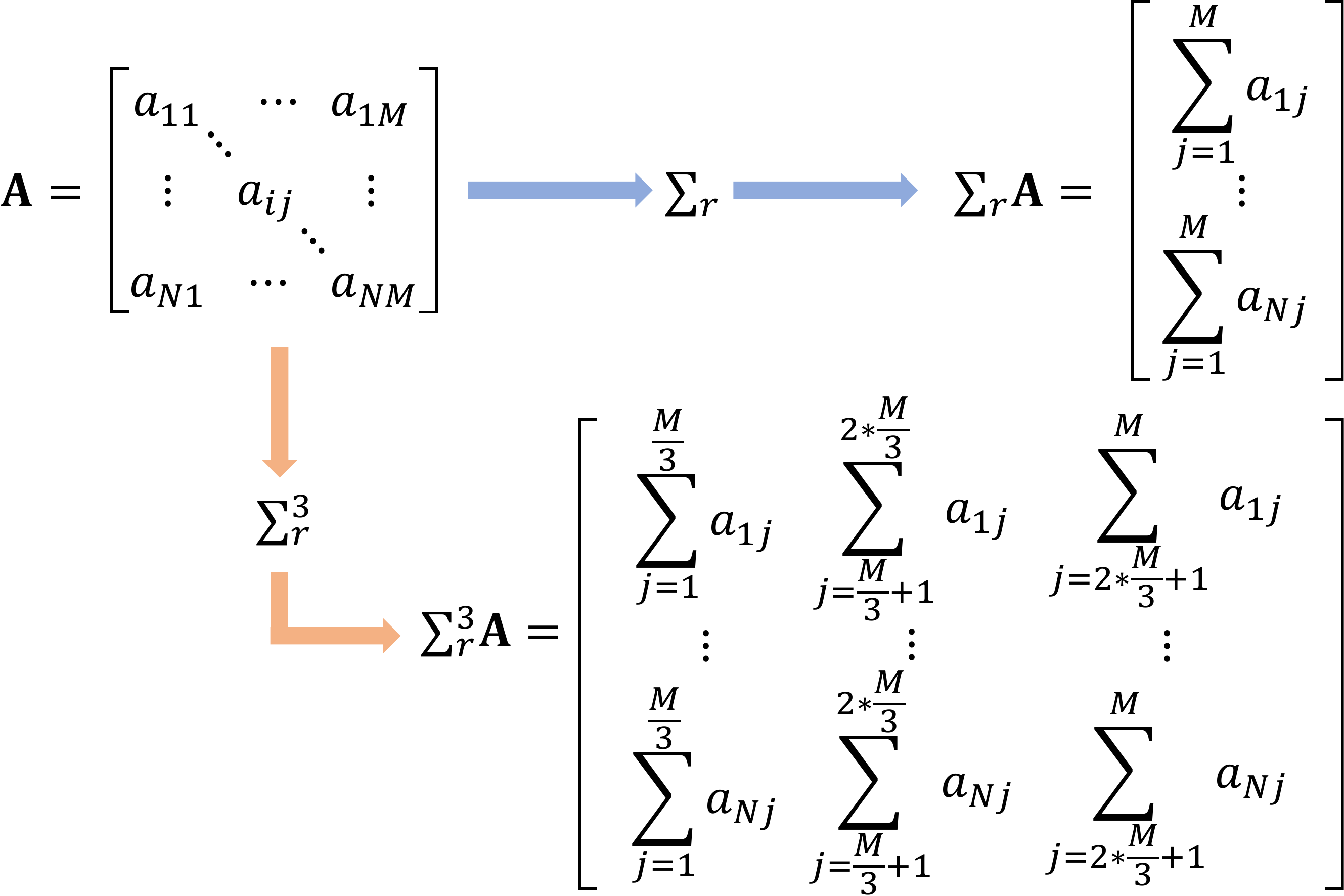}
    \end{center}
    \vspace*{-0.2cm}
    \caption
    {
        Illustration of the row-sum operator $\Sigma_r$ and the trisecting row-sum operator $\Sigma_r^3$. The number $M$ of columns in an operand matrix $\mathbf{A} \in \mathbb{R}^{N \times M}$ is assumed to be a power of 3.
    }
    \label{fig:rowsum_and_rowsum3}
\end{figure}

\begin{algorithm}[b]
\caption{RD-optimized coding (parallel computing)}\label{alg:approach2}
    {\bf Input:} Trit-planes $\mathbf{T}_0, \mathbf{T}_1,\ldots,  \mathbf{T}_{L_{\max}-1}$
    \begin{algorithmic}[1]
        \ForAll{${\mathbf{T}_n}, n=0, \ldots, L_{\max}-1$}
            \State{Set the probability matrix $\mathbf{P}$}
            \State{Set the position matrix $\mathbf{Y}$}
            \State{$\bPy \gets \mathbf{P} \mathbf{Y}$}
            \State{$\bPyy \gets \mathbf{P} \mathbf{Y} \mathbf{Y}$}
            \State{$\mathbf{M} \gets \Sigma_{r} \bPy / \Sigma_{r} \mathbf{P}$}
            \Comment{$\hat{y}^n = E[y|y \in {\cal I}_n]$ in \eqref{eq:recon_of_y}}
            \State{$\mathbf{D}_{n} \gets \Sigma_{r} \bPyy / \Sigma_{r}\mathbf{P} - \mathbf{M}\mathbf{M}$}

            \hspace*{1cm} \Comment{$D_n = E[y^2|y \in {\cal I}_n] - ({\hat{y}^{n}})^2$ in \eqref{eq:distortion}}
            \State{$\mathbf{M} \gets \Sigma_r^3 \bPy / \Sigma_r^3 \mathbf{P}$}
            \State{$\mathbf{D}_{n+1} \gets (\Sigma_r^3 \bPyy - \mathbf{M}\mathbf{M} \Sigma_r^3\mathbf{P}) / \Sigma_{r}\mathbf{P}$}
            \State{$\mathbf{\Delta D} \gets \Sigma_r \mathbf{D}_{n+1} - \mathbf{D}_{n}$}
            \Comment{$\Delta D$ in \eqref{eq:DeltaD}}
            \State{$\mathbf{Q} \gets \Sigma_r^3\mathbf{P} / \Sigma_r\mathbf{P}$}
            \State{$\mathbf{\Delta R} \gets \Sigma_r(-\mathbf{Q}  \log_2{\mathbf{Q}})$}
            \Comment{$\Delta R$ in \eqref{eq:DeltaR}}
            \State{Save the priorities $-\mathbf{\Delta D} / \mathbf{\Delta R}$}
            \State{Sort the trits in the decreasing order of priorities}
            \State{Entropy-encode the sorted trits into the bitstream}
        \EndFor
    \end{algorithmic}
    {\bf Output:} Compressed bitstream
\end{algorithm}

For more efficient implementation, Algorithm \ref{alg:approach2} replaces the nested for-loop in lines 2$\sim$7 in Algorithm \ref{alg:approach1} with matrix operations for parallel computing. In Algorithm \ref{alg:approach2}, $\mathbf{AB}$ and $\mathbf{A}/\mathbf{B}$ denote the Hadamard product and division of matrices $\mathbf{A}$ and $\mathbf{B}$, respectively. Also, the row-sum operator $\Sigma_r$ and the trisecting row-sum operator $\Sigma_r^3$ are illustrated in Figure~\ref{fig:rowsum_and_rowsum3}. Note that a latent element is a continuous Gaussian random variable, but it is centered and quantized into an integer via \eqref{eq:quantization}. Each centered, quantized latent element $y$ belongs to the interval of length $3^{L_{\max}}$. So, for every integer in the interval, we pre-compute the corresponding probability mass of $y$.

Suppose that the $n$th trit-plane ${\mathbf{T}_n}$ is to be encoded using the information in the already encoded $\mathbf{T}_{n-1}, \ldots, \mathbf{T}_{0}$. In other words, given the information that $y$ belongs to ${\cal I}_n$ of length $3^{L_{\max} - n}$, we should encode the trit representing the subinterval ${\cal I}_n^{0}$, ${\cal I}_n^{1}$, or ${\cal I}_n^{2}$ where $y$ belongs. However, depending on the pre-computed PMF of $y$, it may be certain that $y$ belongs to one subinterval, while the other two subintervals have zero probabilities. In such a case, no encoding is necessary. In other words, only `uncertain' trits should be encoded. In line 2 in Algorithm \ref{alg:approach2}, $\mathbf{P} \in \mathbb{R}^{U \times 3^{L_{\max} - n} }$ is the probability matrix, each row of which records the pre-computed probabilities of an uncertain trit in the interval ${\cal I}_n$. Here, $U$ denotes the number of uncertain trits among all $K$ trits. In line 3, $\mathbf{Y} \in \mathbb{R}^{U \times 3^{L_{\max} - n} }$ is the position matrix, each row of which records the integers within the interval ${\cal I}_n$ of $y$. However, it can be shown that the remaining priority computation is invariant to the shift of a latent element. Therefore, each row of $\mathbf{Y}$ is simply filled with $(0, 1, \ldots, 3^{L_{\max} - n}-1)$.

We parallel compute $D_n$ in \eqref{eq:distortion} for every uncertain $y$ in lines 4$\sim$7. Note that $D_n$ can be efficiently computed by
\begin{align}
D_n &= E[(y - \hat{y}^n)^2|y \in {\cal I}_n] \\
    &= E[y^2|y \in {\cal I}_n] - ({\hat{y}^{n}})^2,
\end{align}
so we compute $\bPy$ in line 4 and $\bPyy$ in line 5. Similarly, we parallel compute $\Delta D$ in lines 8$\sim$10 and $\Delta R$ in lines 11$\sim$12. Finally, we compute the priorities in line 13 and perform the RD-optimized coding in line 14$\sim$15.

\begin{figure}[t]
    \begin{center}
    \includegraphics[width=\linewidth]{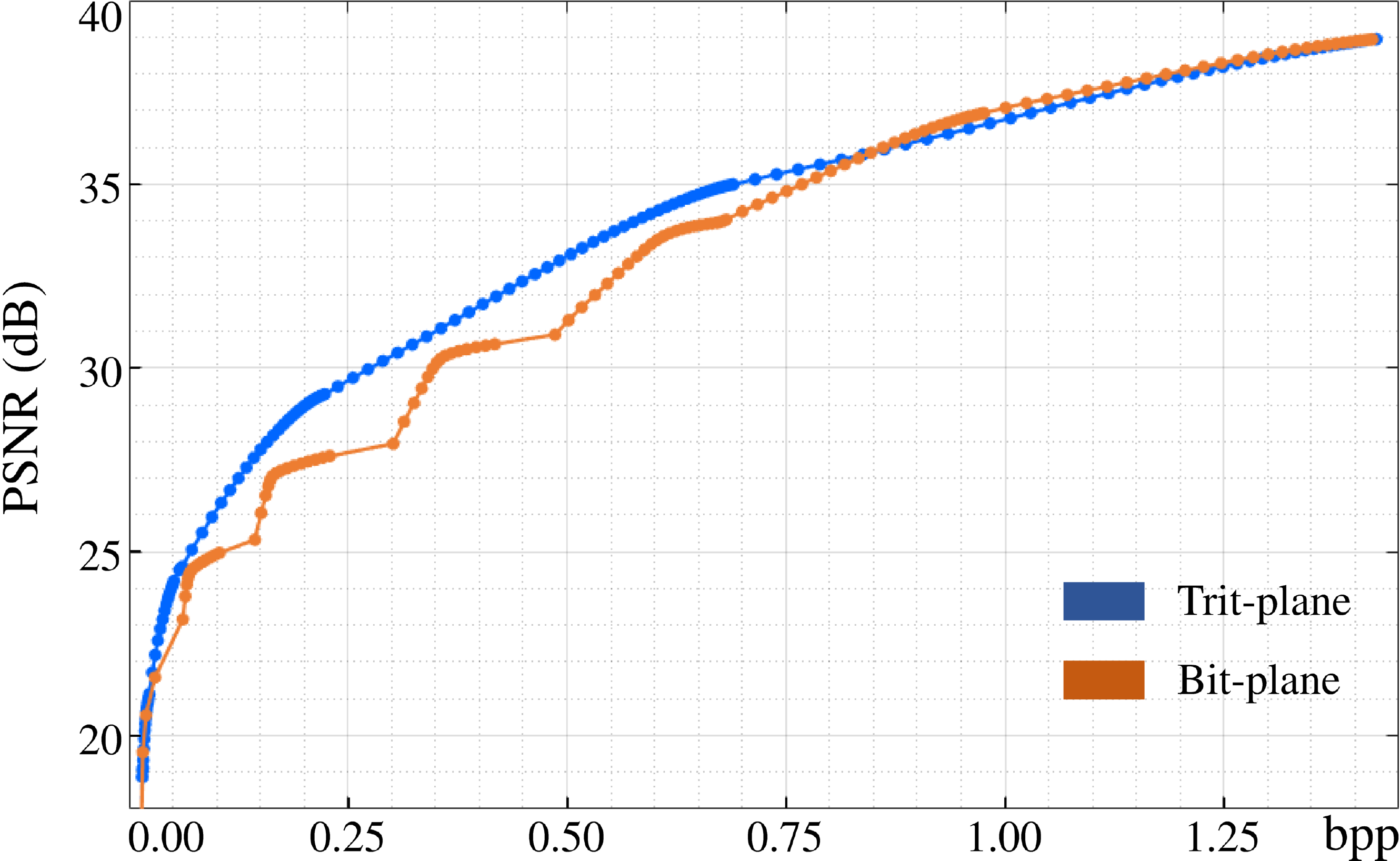}
    \end{center}
    \vspace*{-0.2cm}
    \caption
    {
        RD performance comparison of the trit-plane slicing and the bit-plane slicing.
    }
    \label{fig:rdcurves}
\end{figure}

\vspace*{0.5cm}
\section{Experimental Results}\label{sec:3_experimental_results}
\subsection{Experimental conditions}
We use the Vimeo90k dataset \cite{y2019_IJCV_xue} for training. It contains many frames from overlapping contents, so we sample 80,000 frames only. We crop $256 \times 256$ patches randomly from the training images for data augmentation. We evaluate the compression performance of the proposed DPICT algorithm using the Kodak lossless image dataset \cite{kodim}, composed of 24 images of resolution $512\times768$ or $768\times512$. We use the Adam optimizer\cite{y2015_ICLR_kingma} with a learning rate of $10^{-4}$, a batch size of 16, and $\lambda=0.2$. We train the network for 200 epochs and schedule the learning rate using cosine annealing cycles\cite{y2017_ICLR_huang}. We report a compression bitrate in bits per pixel (bpp), which is obtained by dividing the length of a compressed bitstream by the number of pixels in an image $\mathbf{X}$. We employ PSNR and MS-SSIM \cite{y2003_ACS_wang_MS_SSIM} as distortion metrics. For the arithmetic coder, we use the rANS coder\cite{y2013_arXiv_duda_ANS}.

\begin{figure}[t]
    \begin{center}
    \includegraphics[width=\linewidth]{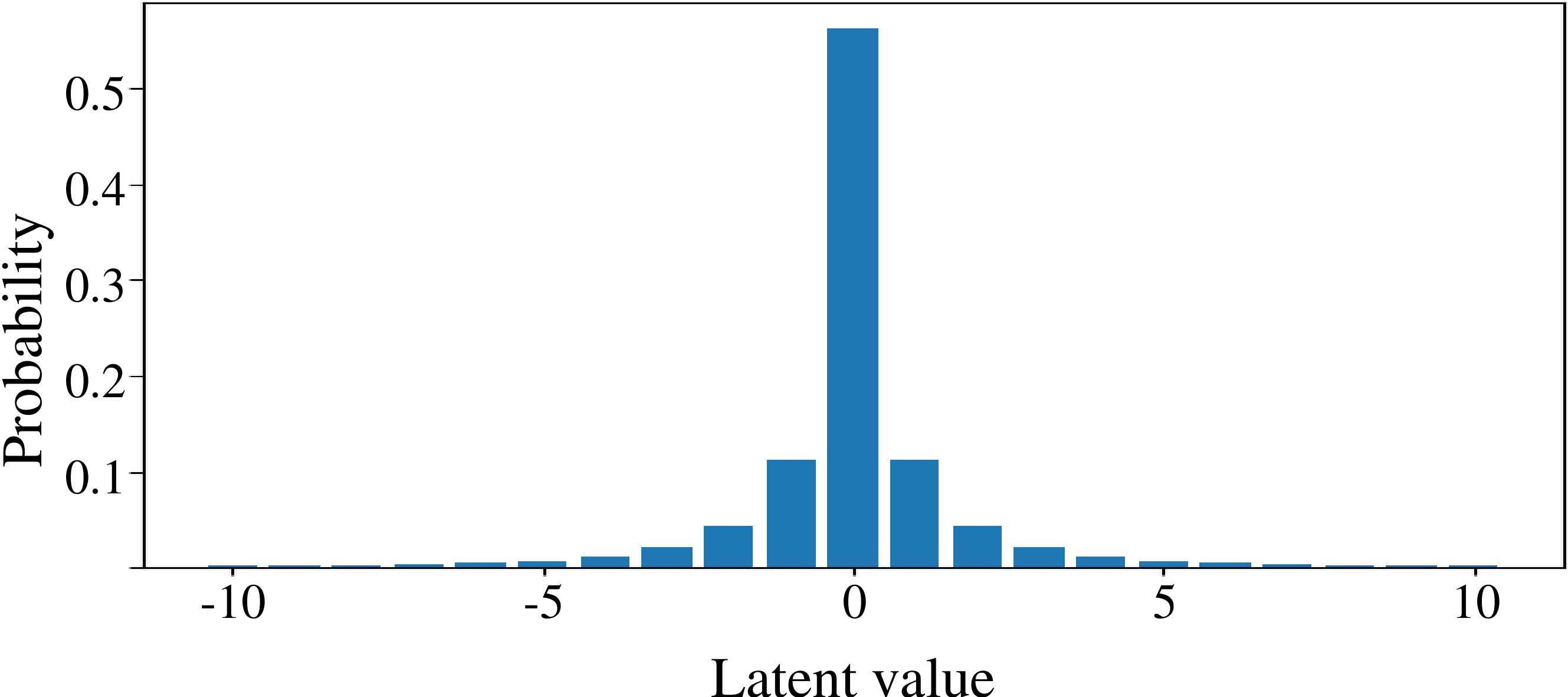}
    \end{center}
    \vspace*{-0.3cm}
    \caption
    {
        The probability distribution of latent values in $\hat{\mathbf{Y}}_c$ for the Kodak dataset.
    }
    \label{fig:histogram}
    \vspace*{0.3cm}
\end{figure}

\begin{figure}[t]
    \begin{center}
    \includegraphics[width=\linewidth]{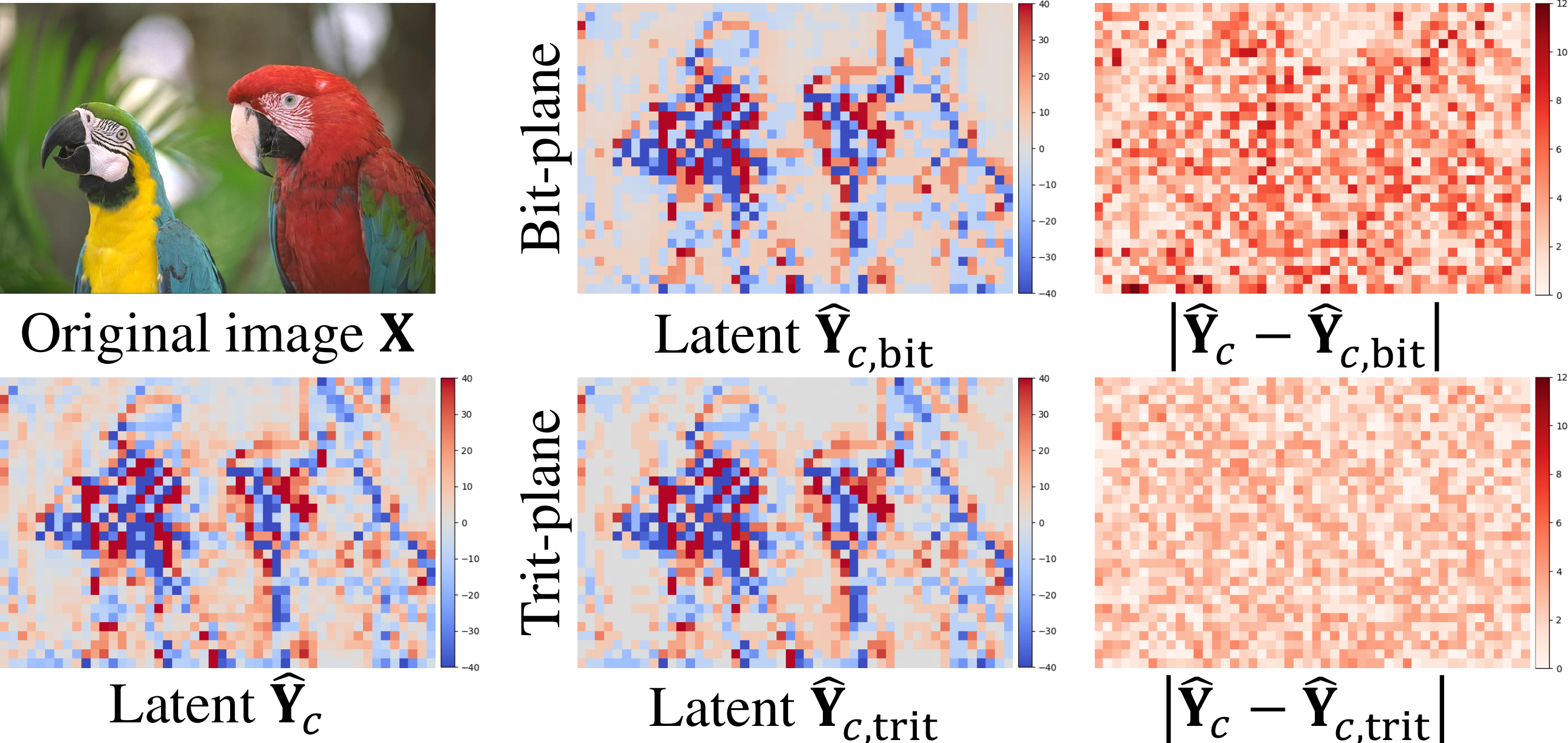}
    \end{center}
    \vspace*{-0.2cm}
    \caption
    {
        Visualization of the latent tensors for the `kodim23' image of the trit-plane slicing and the bit-plane slicing at the same bitrate of 0.22bpp.
    }
    \label{fig:latent}
\end{figure}

\begin{figure*}[t]
    \begin{center}
    \includegraphics[width=\linewidth]{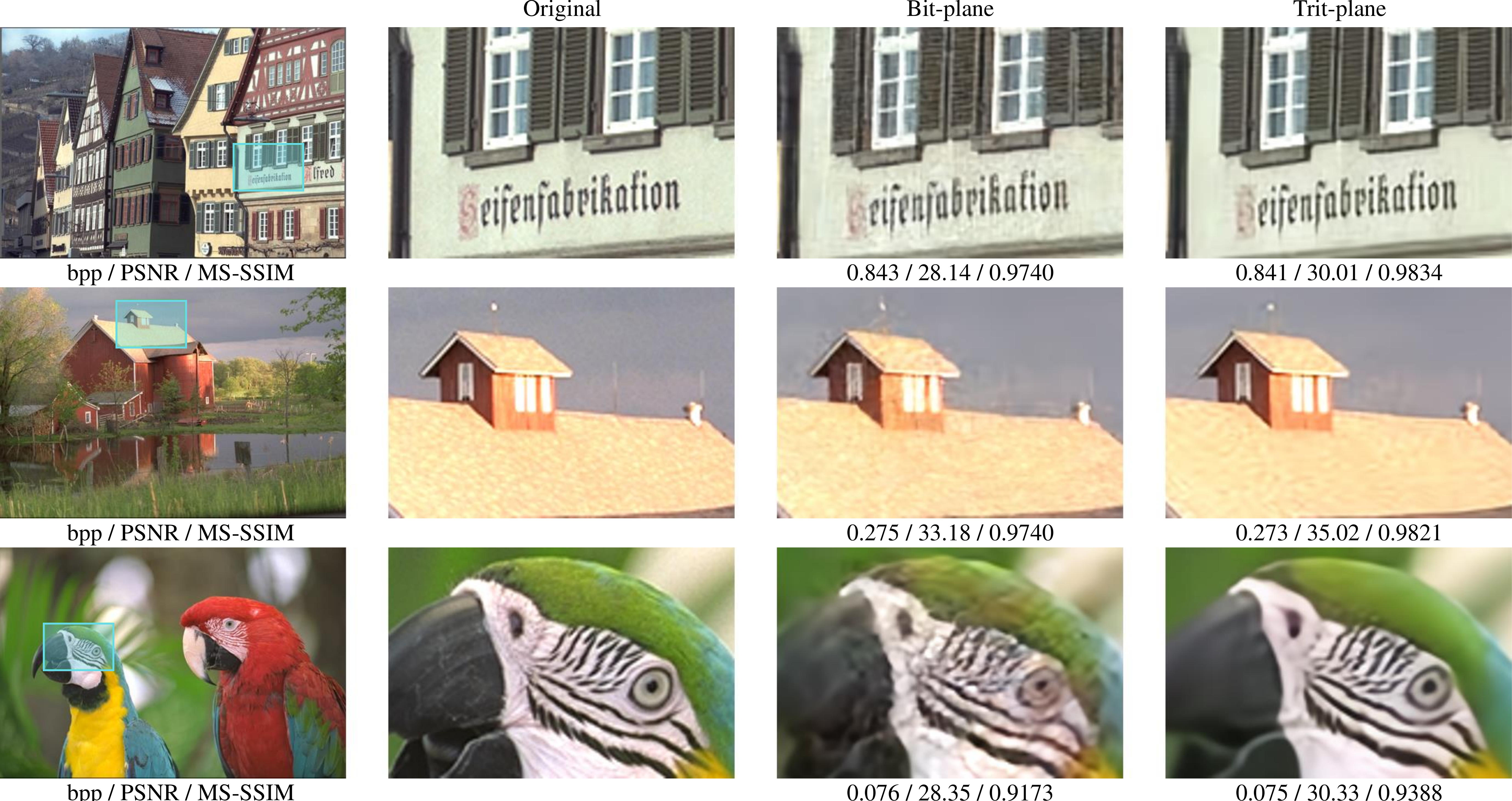}
    \end{center}
    \vspace*{-0.3cm}
    \caption
    {
        Trit-plane slicing vs.~bit-plane slicing: qualitative comparison of reconstructed images at similar rates.
    }
    \label{fig:qualitative}
    \vspace*{0.1cm}
\end{figure*}

\subsection{Performance assessment}
\label{ssec:performance_assessment}
First, we compare the default trit-plane slicing with the alternative bit-plane slicing in Figure~\ref{fig:rdcurves}, which plots the RD curves on the Kodak dataset. The networks are trained with the MSE distortion loss. When the bitrate is lower than 0.9bpp, the trit-plane slicing outperforms the bit-plane slicing. This is because the trit-plane slicing is symmetric around 0 and makes $\hat{y}_i^n$ in \eqref{eq:recon_of_y} zero regardless of $n$ when $y_i$ is zero \cite{y2022_CVPR_DPICT}. In contrast, 0 is not a reconstruction level but a decision level in the bit-plane slicing. Thus, $\hat{y}_i^n$ in \eqref{eq:recon_of_y} becomes nonzero if not all bit-planes are used at the decoder. This degrades reconstructed images because 0 is the most frequent latent value as shown in Figure~\ref{fig:histogram}. On the other hand, at bitrates higher than 0.9bpp, the bit-plane slicing is marginally better than the trit-plane slicing. Note that the second most significant bit-plane corresponds to the quantization interval of 2, whereas the second MST does to that of~3. We conjecture that the finer quantization at high bitrates contributes to marginally better performances of the bit-plane slicing.

Figure~\ref{fig:latent} visualizes the latent tensors for the `kodim23' image in the Kodak dataset, reconstructed by the trit-plane slicing and the bit-plane slicing, respectively, at the same bitrate of 0.22bpp. For visualization, the channel with the maximum entropy is selected, and the latent values are truncated to the interval $[-40, 40]$. In the rightmost column, we compare the reconstruction errors. Because the trit-plane slicing reconstructs original zero values more precisely, it yields less reconstruction errors.

Figure~\ref{fig:qualitative} compares the two slicing schemes qualitatively. At similar bitrates, we see that the trit-plane slicing reconstructs the images --- especially in areas with complicated texture and sharp edges, such as the letters or the facial patterns of the parrot --- more faithfully than the bit-plane slicing does. It also provides higher PSNR and MS-SSIM scores in these examples.

Next, we compare the two implementations of the RD-optimized coding in Algorithms \ref{alg:approach1} and \ref{alg:approach2}. The Kodak test images are encoded and decoded, and the average runtimes are reported in Table~\ref{table:runtime}. As compared to the naive implementation in Algorithm \ref{alg:approach1}, the parallel implementation in Algorithm \ref{alg:approach2} speeds up the encoding and decoding significantly by about 1500 and 600 times, respectively.

\begin{table}[t]
    \caption
    {
        Comparison of Algorithms \ref{alg:approach1} and \ref{alg:approach2}. The average encoding and decoding times (s) on the Kodak dataset are reported.
    }
    \centering
    \small
    \begin{tabular}{l c c}
        \toprule

                & Algorithm \ref{alg:approach1} & Algorithm \ref{alg:approach2}
        \\
        \cmidrule(lr){1-3}

        Encoding & 1703.4 & 1.1
        \\
        Decoding & 1675.0 & 2.8
        \\
        \bottomrule
    \end{tabular}
    \label{table:runtime}
\end{table}

\vspace*{0.5cm}
\section{Conclusions}
In this paper, we developed an efficient implementation of the DPICT algorithm \cite{y2022_CVPR_DPICT}. In DPICT, an input image is transformed into a latent tensor by the encoder network. Then, the latent tensor is represented in the binary or ternary number system and then encoded plane by plane. We analyzed the RD performances of the bit-plane slicing and the trit-plane slicing and demonstrated that the latter --- the default scheme --- provides better RD performances overall. Moreover, to support FGS, DPICT should sort and encode trits in each trit-plane in the decreasing order of RD priorities, which is time-consuming. We developed a parallel computing scheme to speed up the RD-optimized coding. It was shown by experiments that the parallel computing speeds up the encoding and decoding significantly by 1500 and 600 times, respectively.

{\small
\bibliographystyle{ieee_fullname}
\bibliography{2022_arXiv_SMJEON}
}

\end{document}